\documentclass{PoS}

\newcommand{\vs}{\vspace{-2mm}}

\title{Case studies of near-conformal $\beta$-functions}

\ShortTitle{Near-conformal $\beta$-functions}

\author{Zoltan Fodor\\
        University of Wuppertal, Department of Physics, Wuppertal D-42097, Germany\\
        Juelich Supercomputing Center, Forschungszentrum Juelich, Juelich D-52425, Germany\\
        Eotvos University, Pazmany Peter setany 1, 1117 Budapest, Hungary\\
        University of California, San Diego, 9500 Gilman Drive, La Jolla, CA 92093, USA\\
        \email{fodor@bodri.elte.hu}}

\author{\speaker{Kieran Holland}\\
        University of the Pacific, 3601 Pacific Ave, Stockton CA 95211, USA\\
        \email{kholland@pacific.edu}}

\author{Julius Kuti\\
        University of California, San Diego, 9500 Gilman Drive, La Jolla, CA 92093, USA\\
        \email{jkuti@ucsd.edu}}

\author{Daniel Nogradi\\
        Eotvos University, Pazmany Peter setany 1, 1117 Budapest, Hungary\\
        \email{nogradi@bodri.elte.hu}}

\author{Chik Him Wong\\
        University of Wuppertal, Department of Physics, Wuppertal D-42097, Germany\\
        \email{cwong@uni-wuppertal.de}}

\abstract{We present updated results for the non-perturbative $\beta$-function of SU(3) gauge theories with $N_f = 12$ or 10 massless flavors in the fundamental rep or $N_f = 2$ in the sextet rep, measured with staggered fermions. New data at finer lattice spacing and our previously introduced method, the infinitesimal $\beta$-function, strengthen the case that the $N_f = 12$ model has no infrared fixed point up to $g^2 = 7.2$. We show how underestimated cutoff dependence in one domain wall study for $N_f = 10$ has been corrected, which is now consistent with staggered results showing a monotonically increasing $\beta$-function. A consistent theme is that too small volumes can lead to apparent fixed points which vanish towards the continuum limit. We also apply the infinitesimal $\beta$-function method to the $N_f = 10$ model, finding consistent behavior with the finite-step $\beta$-function. Ongoing simulations and analysis for the sextet model confirm our previous results from weak to strong coupling with a non-zero $\beta$-function throughout, in quantitative difference to Wilson fermion simulations~\cite{Hasenfratz:2015ssa}.}

\FullConference{The 37th Annual International Symposium on Lattice Field Theory - LATTICE2019\\
		16-22 June, 2019\\
		Wuhan, China.}

\begin{document}

\section{$N_f = 12$}

\begin{figure}[h]
\begin{center}
     \includegraphics[width=.48\textwidth]{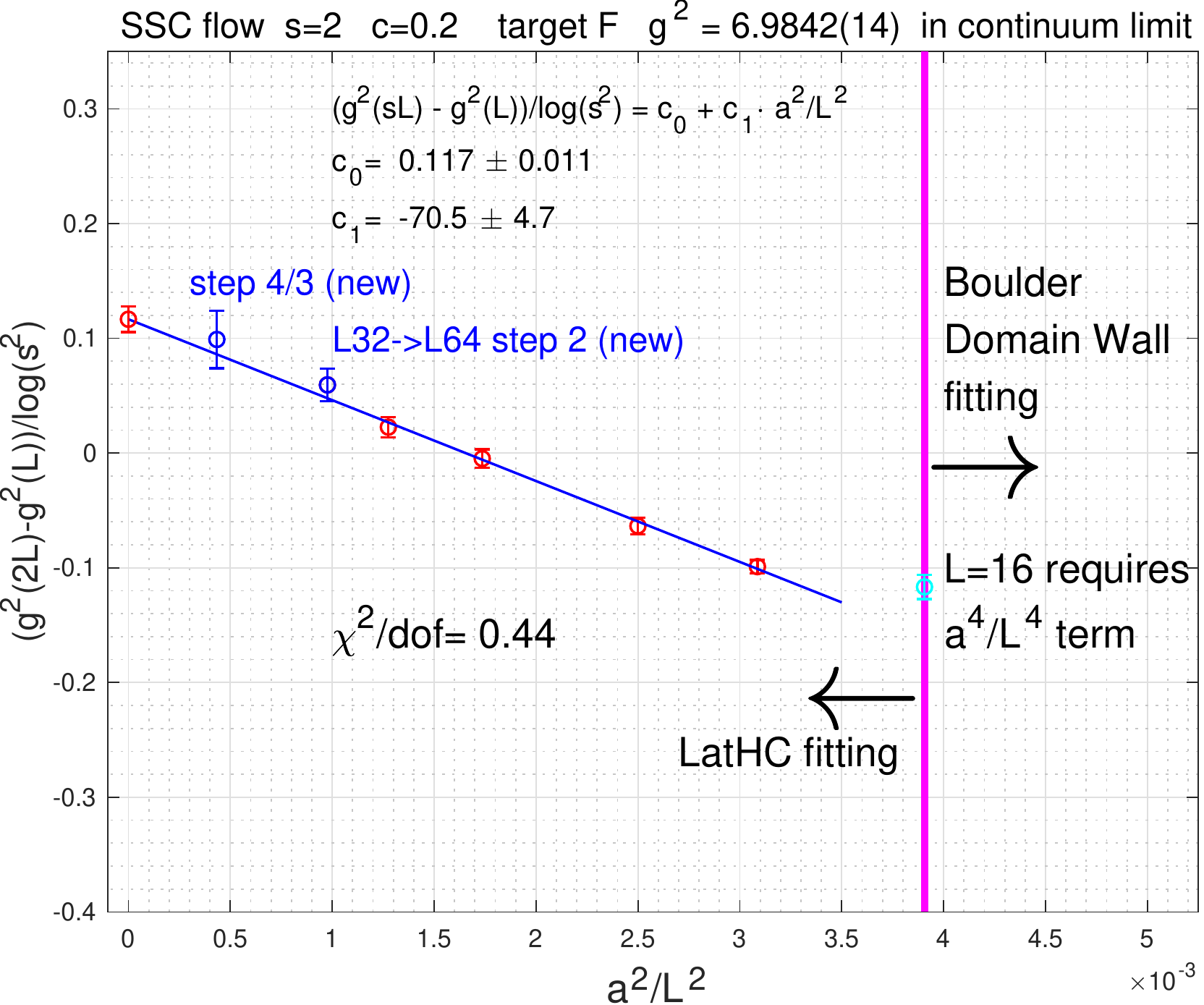}
     \includegraphics[width=.48\textwidth]{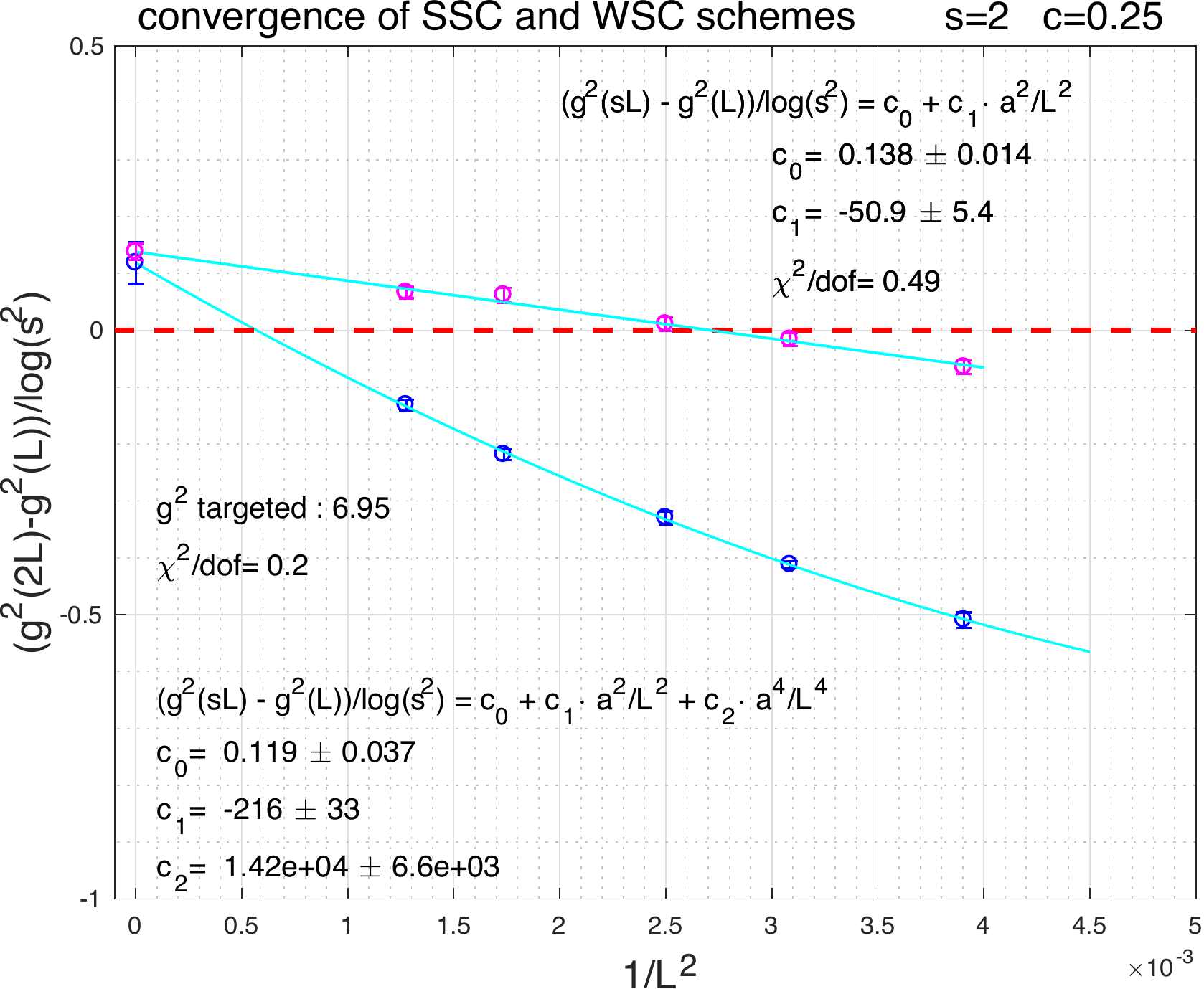}
     \caption{(Left) Continuum extrapolation of the step-scaling $\beta$-function for $N_f = 12$. The new finest lattice spacing data point $32 \rightarrow 64$ is fully consistent with previous results, with the overlaid point $48 \rightarrow 64$ with $s = 4/3$ being very close to the continuum value for $s = 2$. (Right) Consistent continuum limits for Symanzik (magenta) and Wilson (blue) discretizations of the gradient flow.}
     \label{fig1}
\end{center}
\end{figure}

Lattice simulations are the tool of choice for non-perturbative determination of gauge theory $\beta$-functions. This is particularly relevant for exploring models with possible near-conformal behavior. The use of the gradient flow has become standard, with step-scaling for continuum extrapolation. There have been repeated claims of an infrared fixed point (IRFP) for the SU(3) $N_f = 12$ fundamental rep $\beta$-function~\cite{Cheng:2014jba, Hasenfratz:2016dou}, of late using domain wall fermions~\cite{Hasenfratz:2017mdh, Hasenfratz:2017qyr, Hasenfratz:2019dpr}, inconsistent with our staggered studies~\cite{Fodor:2017gtj}. To probe further the disagreement, we have generated new tuned data for $c = \sqrt{8t}/L = 0.2$ at finer lattice spacing $32 \rightarrow 64$, which we find are completely in line with previous data and analysis (see Fig.\ref{fig1}), reaffirming the absence of a fixed point in the range $0 < g^2 < 7.2$. Comparison of Symanzik and Wilson variants of interpolated gradient flow data at $c = 0.25$ give compatible continuum results, with the Symanzik $\beta$-function already positive at the finest lattice spacings. The $\beta$-function is $s$-dependent, however overlaying the data point $48 \rightarrow 64$ for $s = 4/3$ emphasizes the proximity to $a/L = 0$, in contrast to domain wall simulations with $16 \rightarrow 32$ or coarser, for which ${\cal O}(a^4)$ fitting appears necessary.

\begin{figure}[h]
\begin{center}
     \includegraphics[width=.48\textwidth]{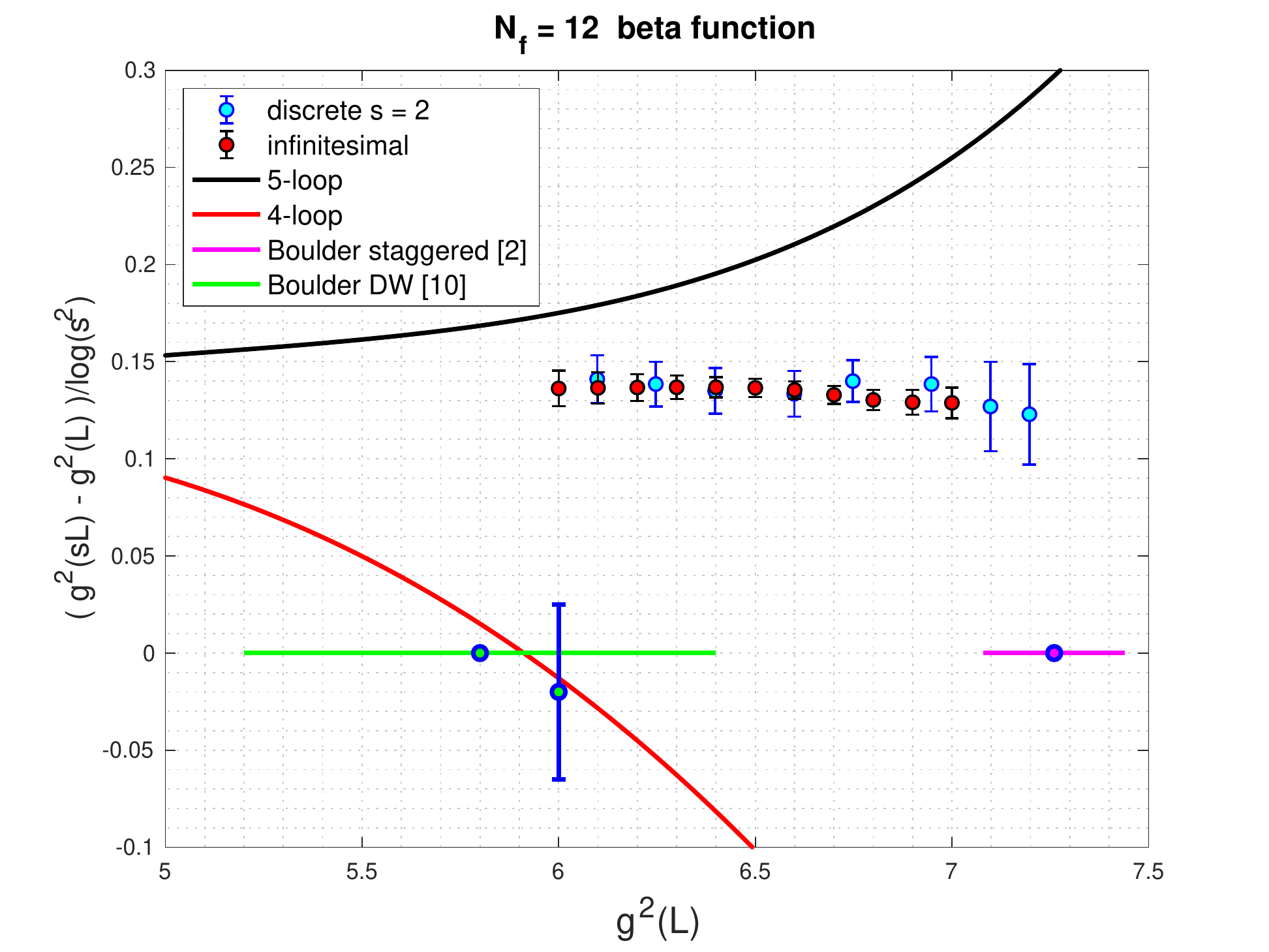}
     \includegraphics[width=.43\textwidth]{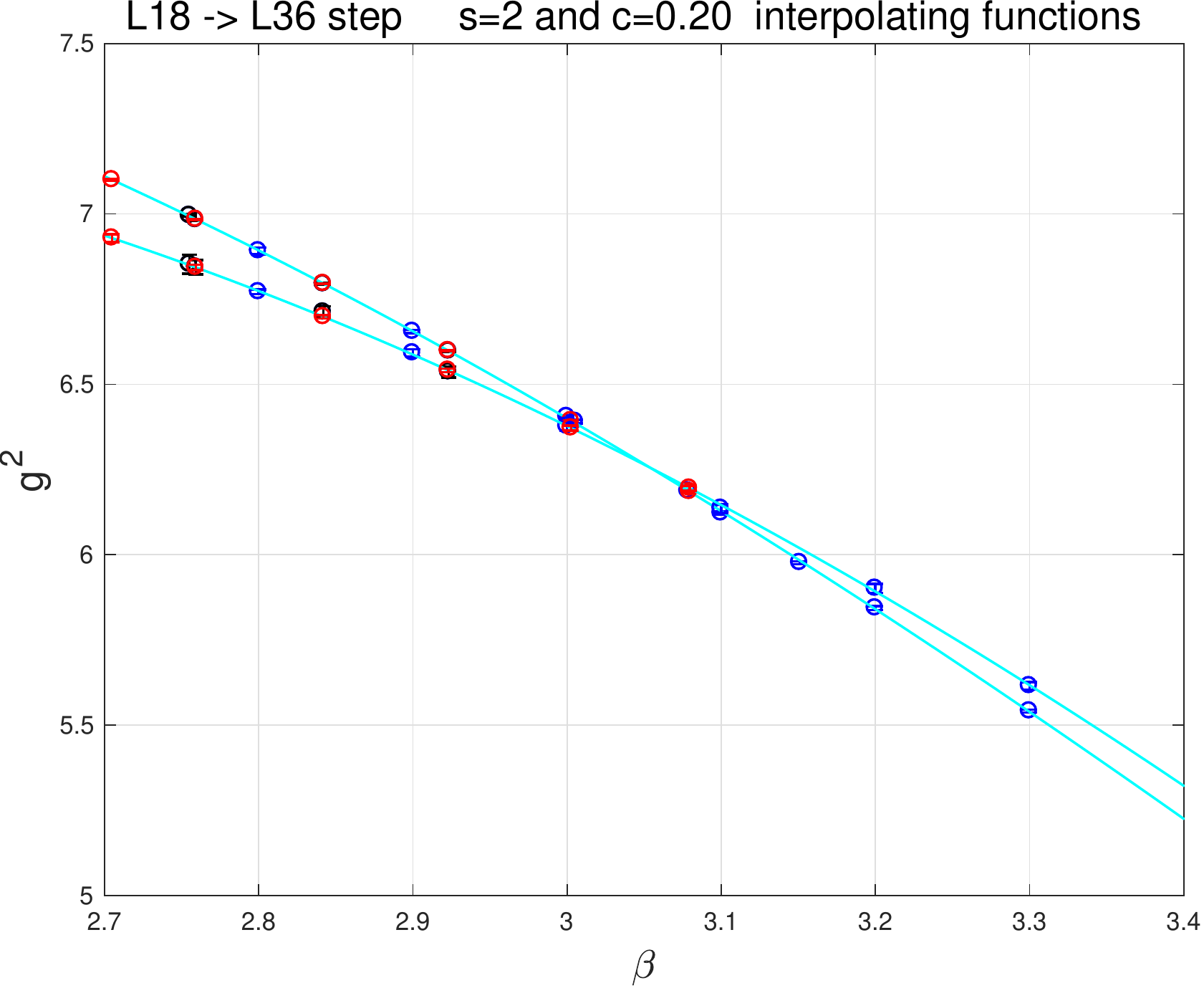}
     \caption{(Left) Excellent agreement between the finite-step $s=2$ $c = 0.25$ and infinitesimal $\beta$-functions for $N_f = 12$, essentially constant across the explored $g^2$ range. Also shown are IRFP predictions from~\cite{Hasenfratz:2016dou} with staggered and~\cite{Hasenfratz:2019dpr} with domain wall, with an example of the attained accuracy. (Right) An example for $N_f = 12$ of an apparent fixed point at $g^2 \sim 6.3$ at coarse lattice spacing with paired volumes $18 \rightarrow 36$. Blue points are the input data and red points are interpolations to targeted $g^2$ values.}
     \label{fig2}
\end{center}
\end{figure}
Our continuum $N_f = 12$ $\beta$-function is flat in the region $6.0 \leq g^2 \leq 7.2$, covering the range where IRFPs have previously been claimed (see Fig.\ref{fig2}). At e.g.~$g^2 = 6.98$ the statistical significance of our result is more than $10 \sigma$ away from zero and incompatible with a fixed point. Comparing the values $c = 0.20$ and 0.25 defining the finite-volume renormalization scheme, we find the same qualitative behavior with no IRFP appearing, the value of $c$ is not important. In addition we use our previously introduced method~\cite{Fodor:2017die} to calculate the infinitesimal $\beta$-function $t \cdot dg^2/dt$ directly from lattice simulations (described in Sec.4), we show in Fig.\ref{fig2} a summary of the results with excellent agreement with the finite-step $\beta$-function. How a transient fixed point can appear at coarse lattice spacing is shown in Fig.\ref{fig2} for paired volumes $18 \rightarrow 36$, where $g^2$ as a function of the bare coupling crosses at $g^2 \sim 6.3$. Pushing to finer lattice spacing as in Fig.\ref{fig3} with volumes $32 \rightarrow 64$ the fixed point has disappeared with a positive $\beta$-function throughout. The initial claim of an IRFP at $g^2 \sim 6$ with staggered fermions~\cite{Cheng:2014jba} was moved to $g^2 \sim 7$~\cite{Hasenfratz:2016dou}, all based on smaller volume simulations, and was not reconciled with the larger volume results of no IRFP for our staggered studies. The IRFP is moved back to $g^2 \sim 6$ with numerically-expensive domain wall fermions~\cite{Hasenfratz:2019dpr} which cannot reach the larger volumes and accuracy needed to resolve the existence or not of a continuum IRFP e.g.~with $c = 0.25$ the quoted result is $-0.065 \lesssim \beta(g^2=6) \lesssim 0.025$.

\vspace{-5mm}
\section{$N_f = 10$}
\begin{figure}[h]
\begin{center}
     \includegraphics[width=.43\textwidth]{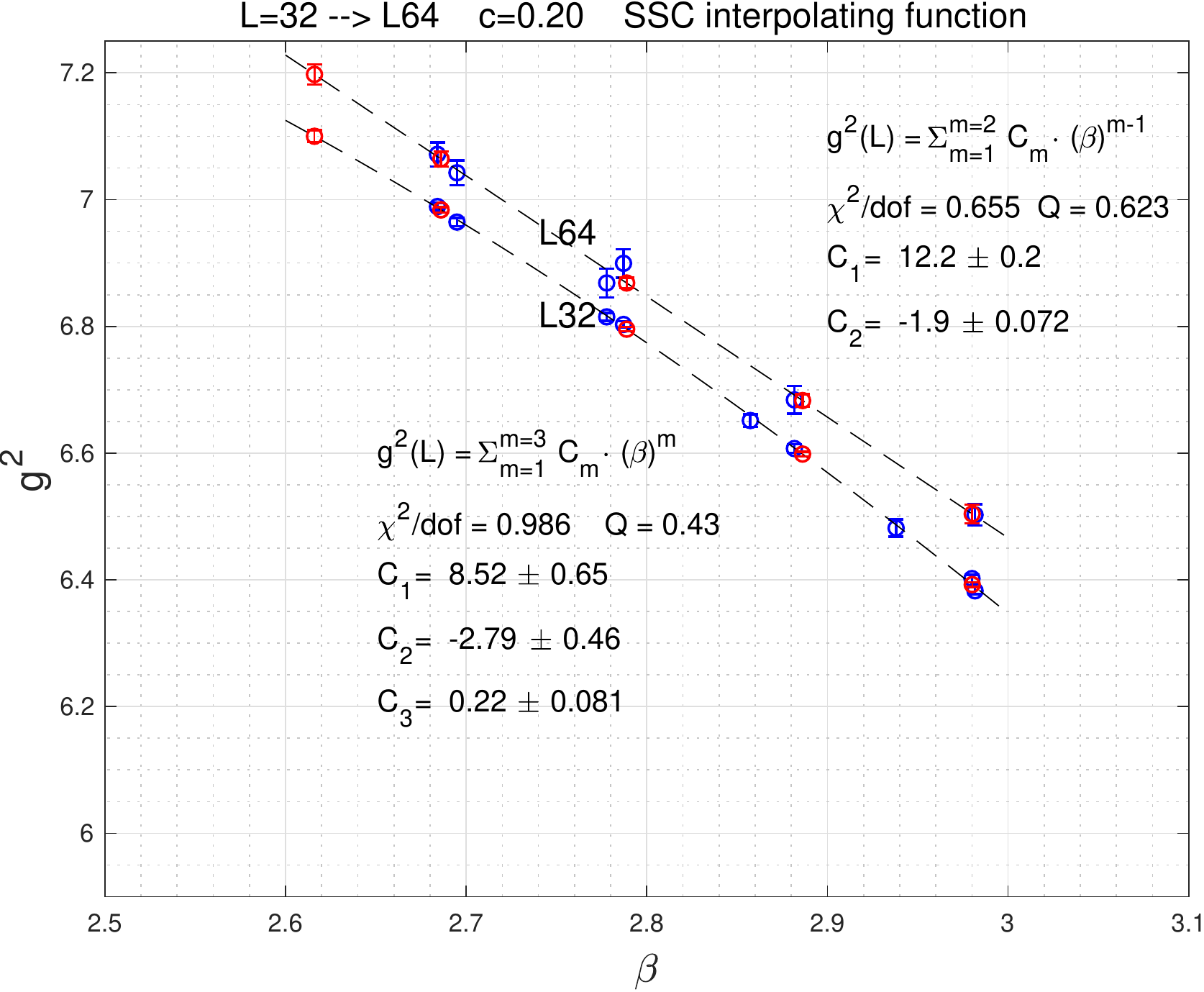}
     \includegraphics[width=.48\textwidth]{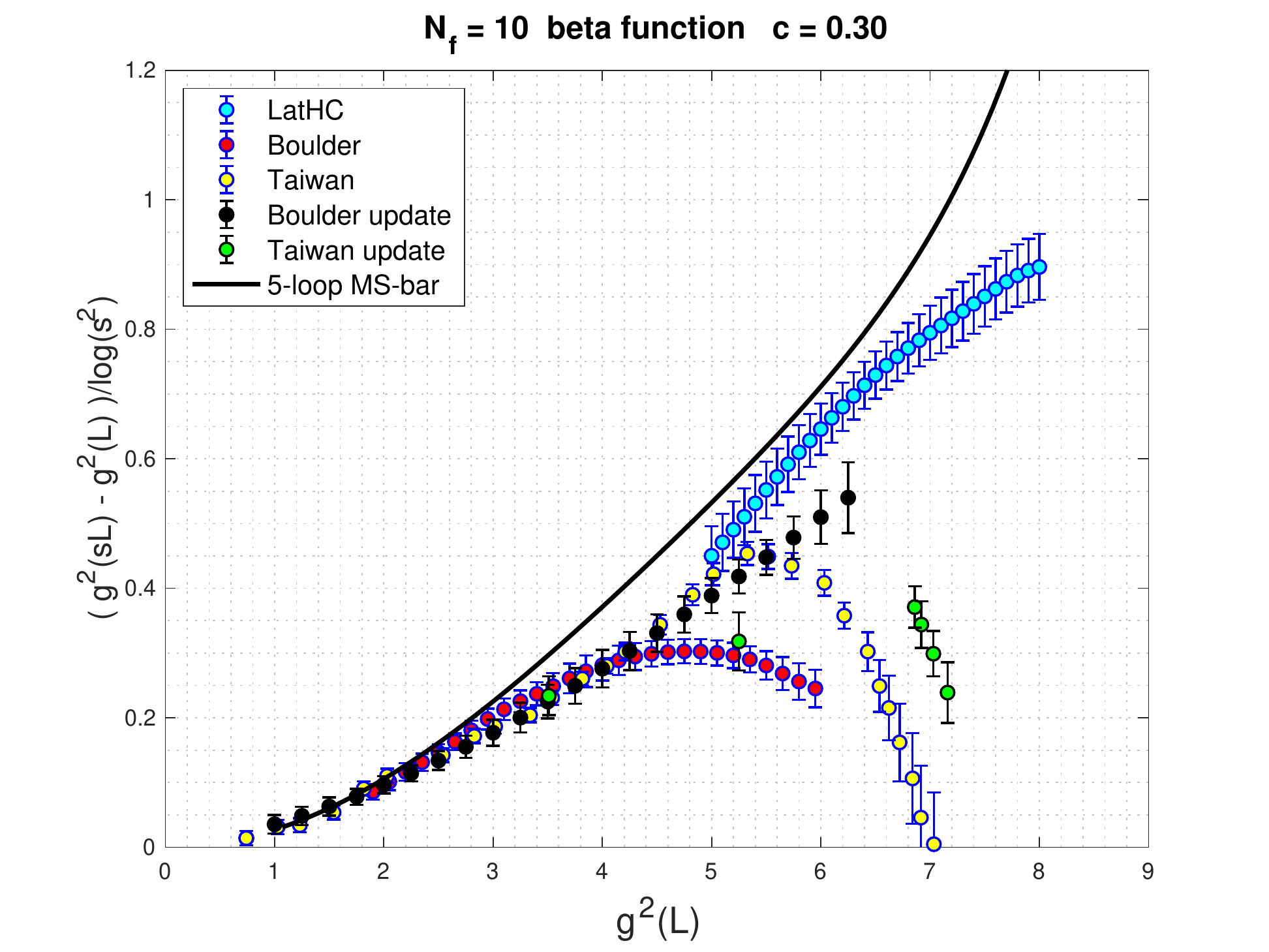}
     \caption{(Left) The disappearance of the $N_f = 12$ fixed point in Fig.\ref{fig2} at finer lattice spacing for volumes $32 \rightarrow 64$, with the discrete-step $\beta$-function positive throughout. (Right) Comparison of recent staggered and domain wall determinations of the $N_f = 10$ $\beta$-function.}
     \label{fig3}
\end{center}
\end{figure}

\begin{figure}[h]
  \begin{center}
         \includegraphics[width=.48\textwidth]{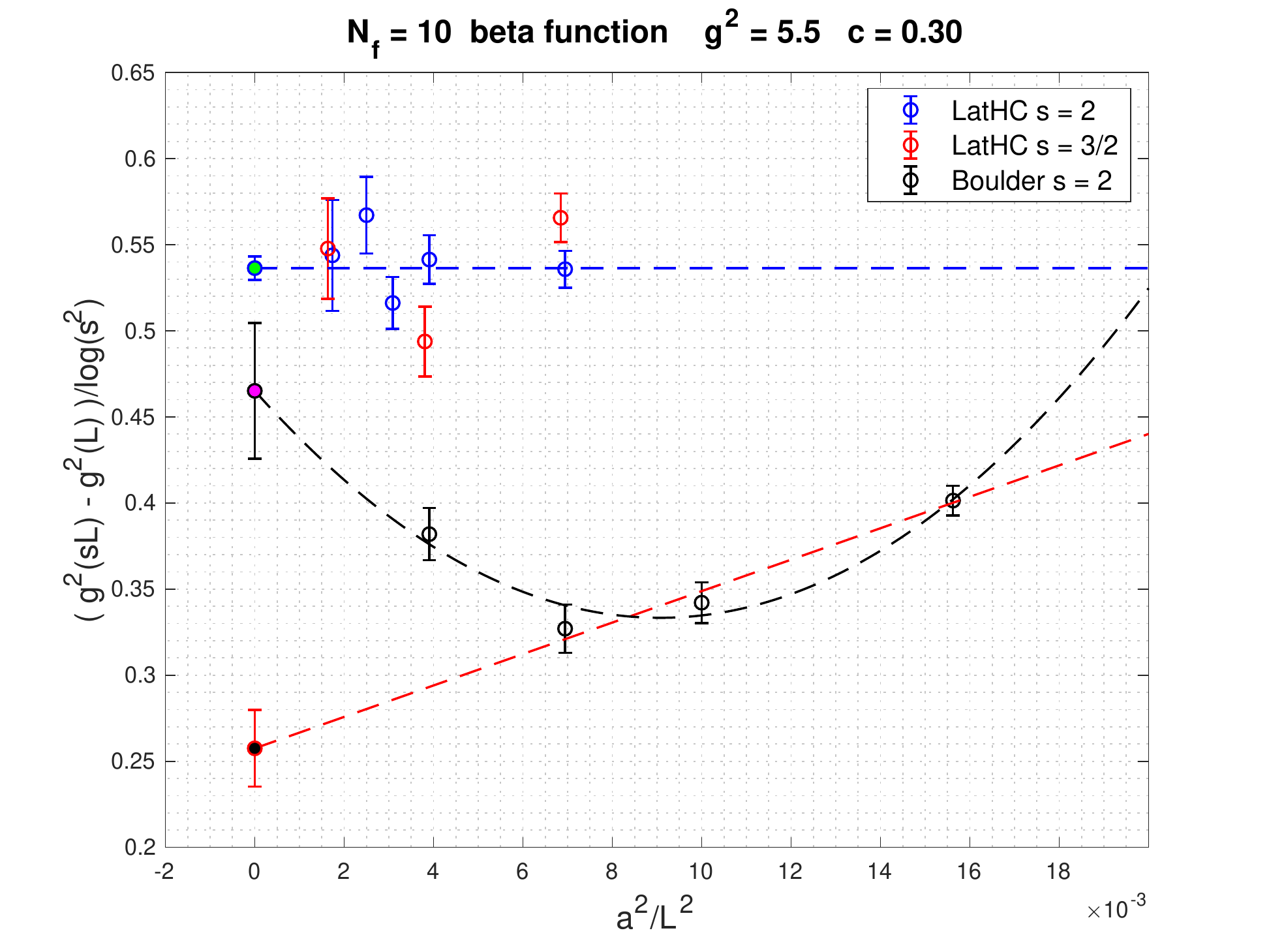}
     \includegraphics[width=.48\textwidth]{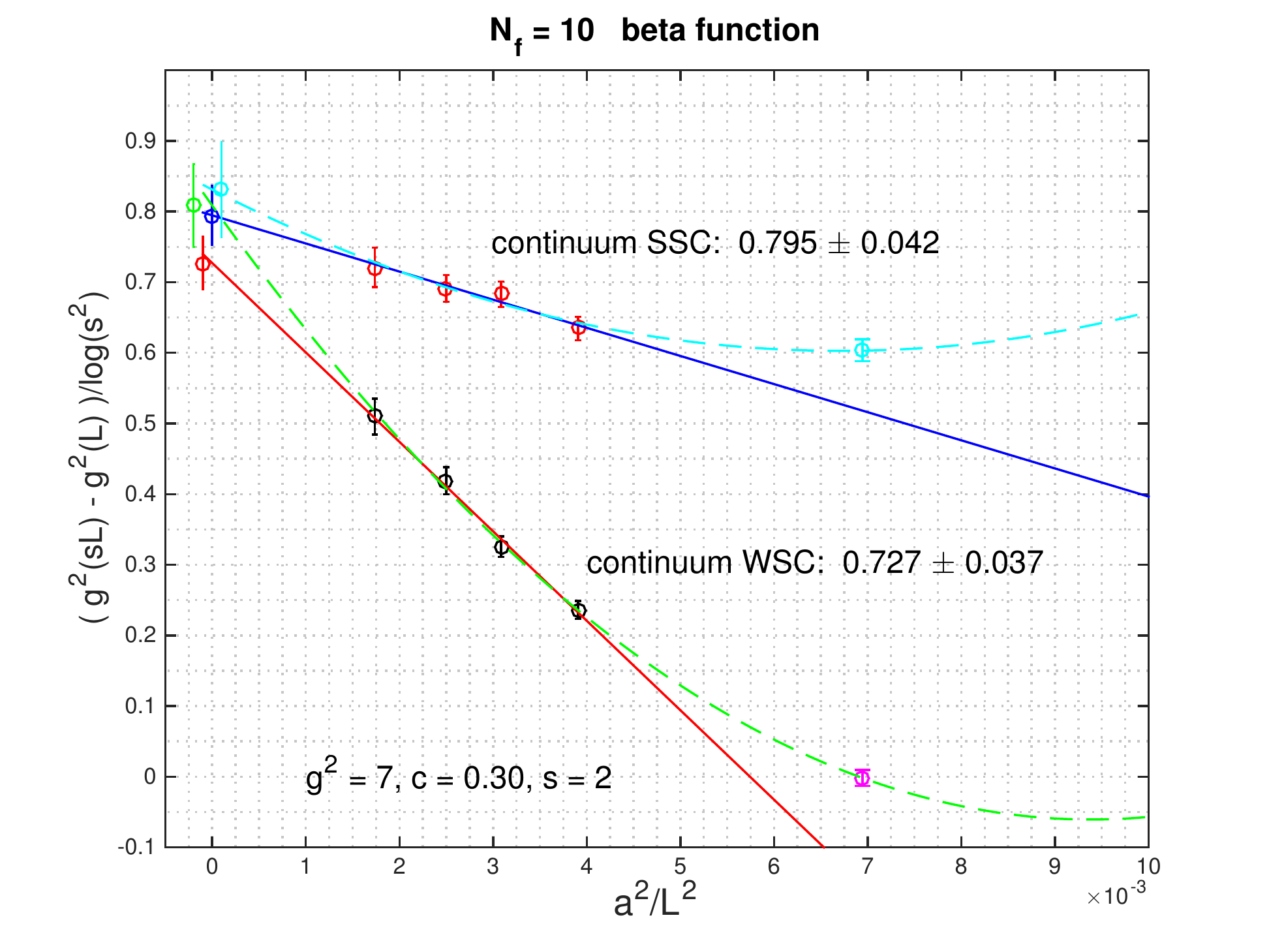}
     \caption{(Left) Comparison of continuum extrapolations for staggered (LatHC) and domain wall (Boulder). The new domain wall data point $L/a = 16$ shows ${\cal O}(a^4)$ dependence, giving a compatible result with staggered, resolving the discrepancy given by ${\cal O}(a^2)$ fitting of domain wall with $L/a = 8, 10, 12$. (Right) Systematic tests for $N_f = 10$, varying the gradient flow and comparing ${\cal O}(a^2)$ and ${\cal O}(a^4)$ extrapolations where $L/a = 12$ is omitted or included.}
     \label{fig5}
\end{center}
\end{figure}

SU(3) gauge theory with $N_f = 10$ massless fundamental rep flavors is an important anchor point: if conformal with an IRFP, so also is the $N_f = 12$ model with an IRFP at weaker coupling. The $N_f = 10$ model is also a template for the Higgs as a pseudo-Goldstone boson.
A summary of recent lattice work on the $\beta$-function of SU(3) with $N_f = 10$ flavors is in Fig.\ref{fig3}. Staggered studies show the $\beta$-function increases monotonically in qualitative agreement with 5-loop perturbative calculations~\cite{Fodor:2018tdg}. Initial domain wall studies disagreed with one another at stronger coupling, with one~\cite{Chiu:2017kza} predicting an IRFP at $g^2 \sim 7$, and the second showing a smaller downturn~\cite{Hasenfratz:2017mdh}. Both domain wall results have been significantly changed much beyond statistical errors, one with the IRFP no longer evident~\cite{Chiu:2018edw} and the other~\cite{Hasenfratz:2017qyr} now compatible with the staggered measurement, with no indication of an IRFP. The cause is clear in Fig.\ref{fig5} (left). Step scaling $L \rightarrow sL$ with $L/a = 8, 10, 12$ shows apparent ${\cal O}(a^2)$ lattice artifacts for the domain wall calculation, but additional data at finer lattice spacing $L/a = 16$ exposes the necessary ${\cal O}(a^4)$ term which dramatically increases the continuum domain wall result. The concern is shared with $N_f = 12$ studies: accurate data on small volumes at coarse lattice spacing may show spurious ${\cal O}(a^2)$ scaling, leading to an underestimate of the $\beta$-function which only comes to light when larger volumes are added. Staggered data with finer lattices up to $L/a = 24$ typically show ${\cal O}(a^2)$ scaling, at this coupling the artifacts happen to be small, the $\beta$-function is $s$-dependent but $s = 2$ and $3/2$ data give very similar continuum results.  

\begin{figure}[h]
\begin{center}
     \includegraphics[width=.48\textwidth]{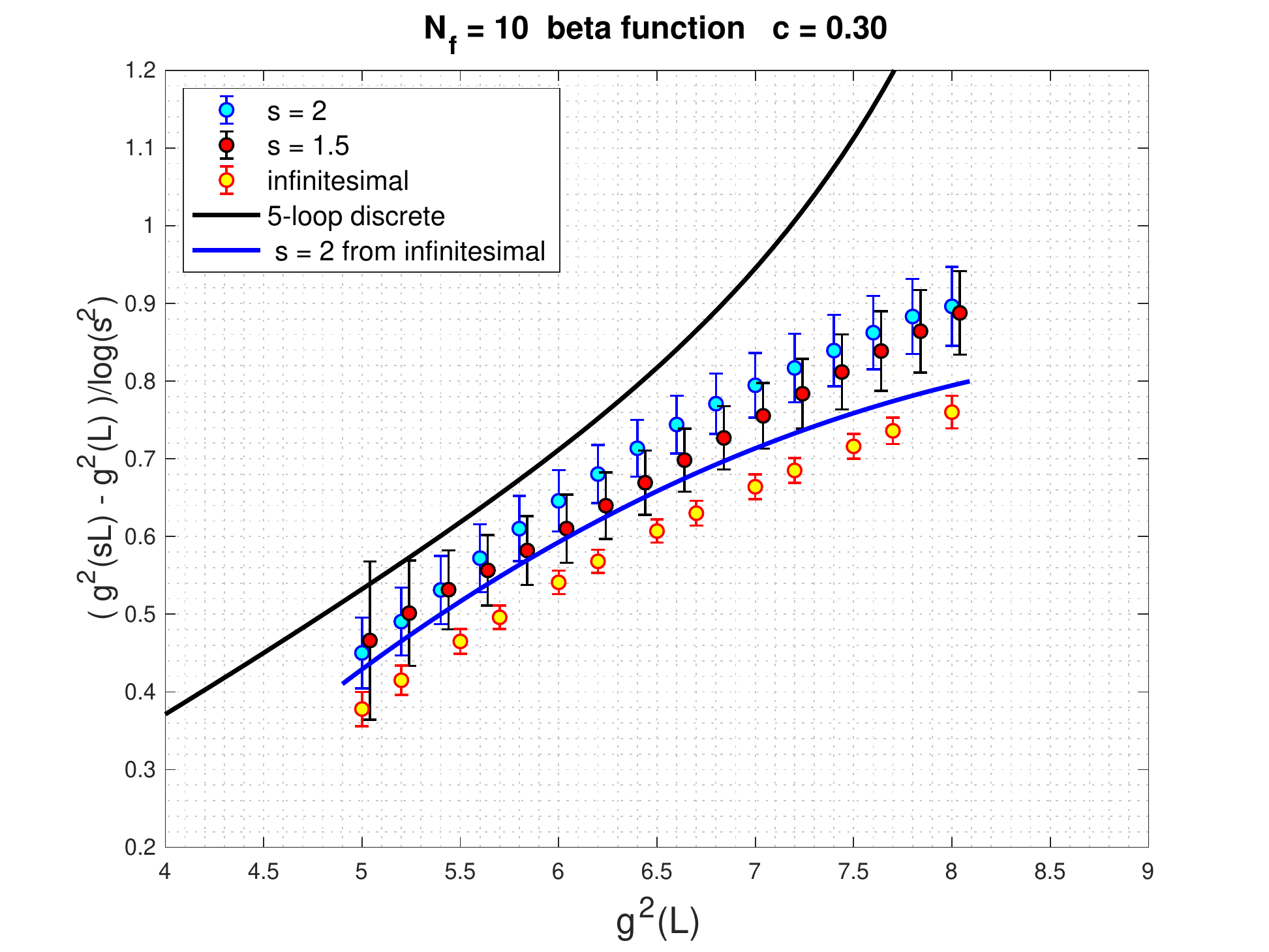}
     \includegraphics[width=.48\textwidth]{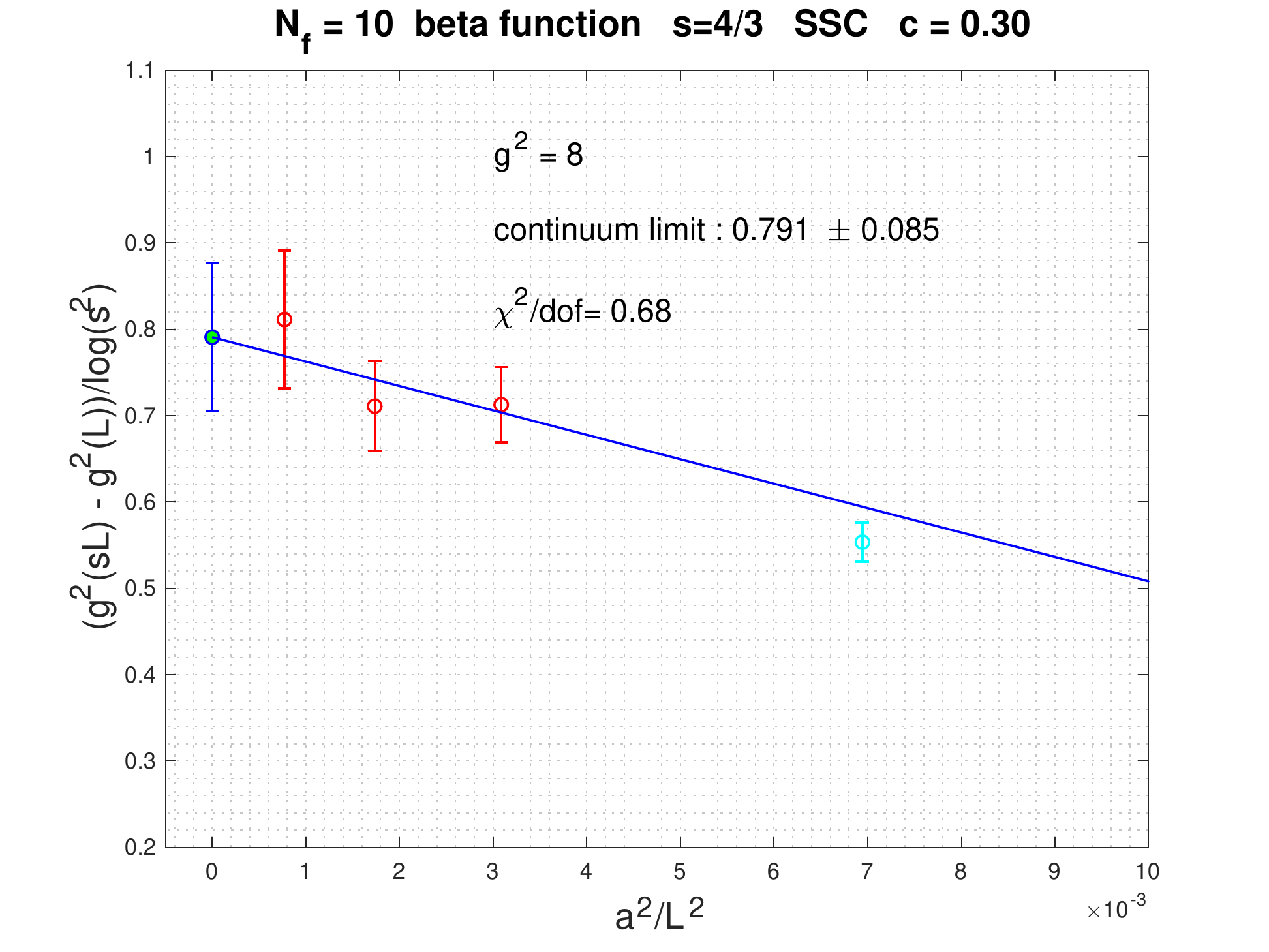}
     \caption{(Left) The $N_f = 10$ finite-step $\beta$-function shows mild dependence on the step scale $s$ value. The infinitesimal $\beta$-function shows the same behavior, and by integration can be used to predict the $s=2$ finite-step function. (Right) The choice $s = 4/3$ for $N_f = 10$ yields a finest lattice spacing $36 \rightarrow 48$ close to the continuum limit, and an extrapolated value far above zero. }
     \label{fig9}
\end{center}
\end{figure}

We test systematically various discretizations of the gradient flow coupling. The Symanzik gauge action is used in MC simulation, for the gradient flow we use both Symanzik and Wilson variants, as well as Symanzik or Clover versions of the observable. We show in Fig.\ref{fig5} nice continuum agreement between two flow variants at a targeted strong coupling $g^2 = 7$. Inclusion of $L/a = 12$ not surprisingly requires an ${\cal O}(a^4)$ term in the continuum extrapolation and is fully consistent with an ${\cal O}(a^2)$ fit of data with $L/a \ge 16$. We find Symanzik flow and Clover observable i.e.~SSC has the smallest cutoff effects and is our natural reference choice. 
We explore various choices for the step-scale $s$ and the finite-volume renormalization scheme $c = \sqrt{8t}/L$ defining the $\beta$-function. We compare in Fig.\ref{fig9} two values $s = 2$ and $3/2$ and find only mild variation. Also included is our new infinitesimal $\beta$-function determination, showing the same qualitative behavior. Given our set of volumes, the choice $s = 4/3$ allows us to push closer to the continuum limit with the finest lattice spacing corresponding to $36 \rightarrow 48$. With mild cutoff effects $\sim 10$\% for the three finer lattices, it is not possible to reconcile with the much smaller result of~\cite{Chiu:2018edw}. An additional test (not shown) is a comparison of continuum results for $c = 0.30$ and 0.25. The former value, used in domain wall studies, is important for direct comparison, however the $\beta$-function can be equally well-determined at the smaller $c$ value, where lattice artifacts can be well fit by an ${\cal O}(a^2)$ form. The broad conclusions are the same: at other $c$ values we find monotonic increase in the $\beta$-function in the range $5 \leq g^2 \leq 8$.

\vspace{-10mm}
\section{$N_f = 2$ sextet}

\begin{figure}[h]
\begin{center}
     \includegraphics[width=.48\textwidth]{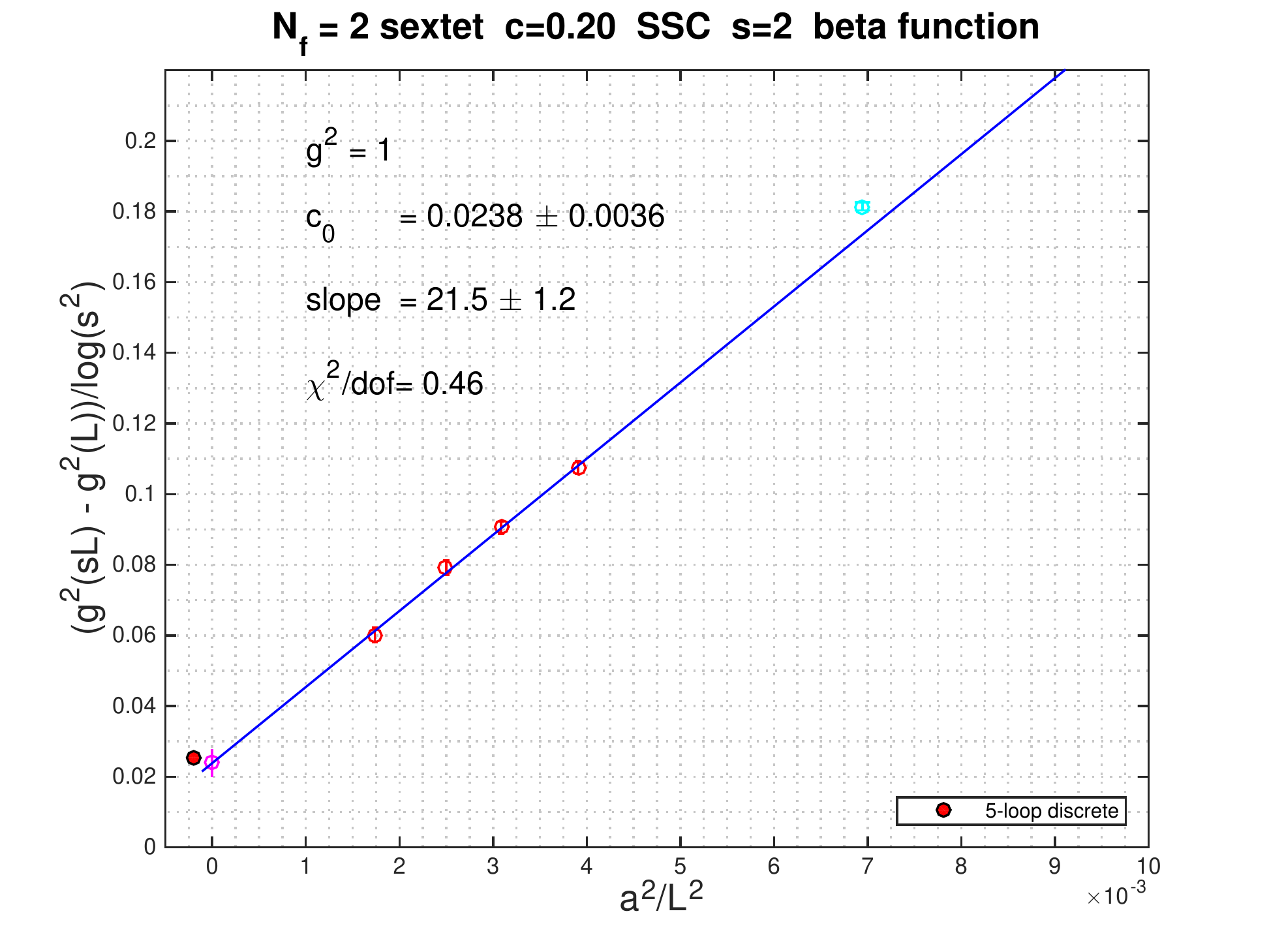}
     \includegraphics[width=.48\textwidth]{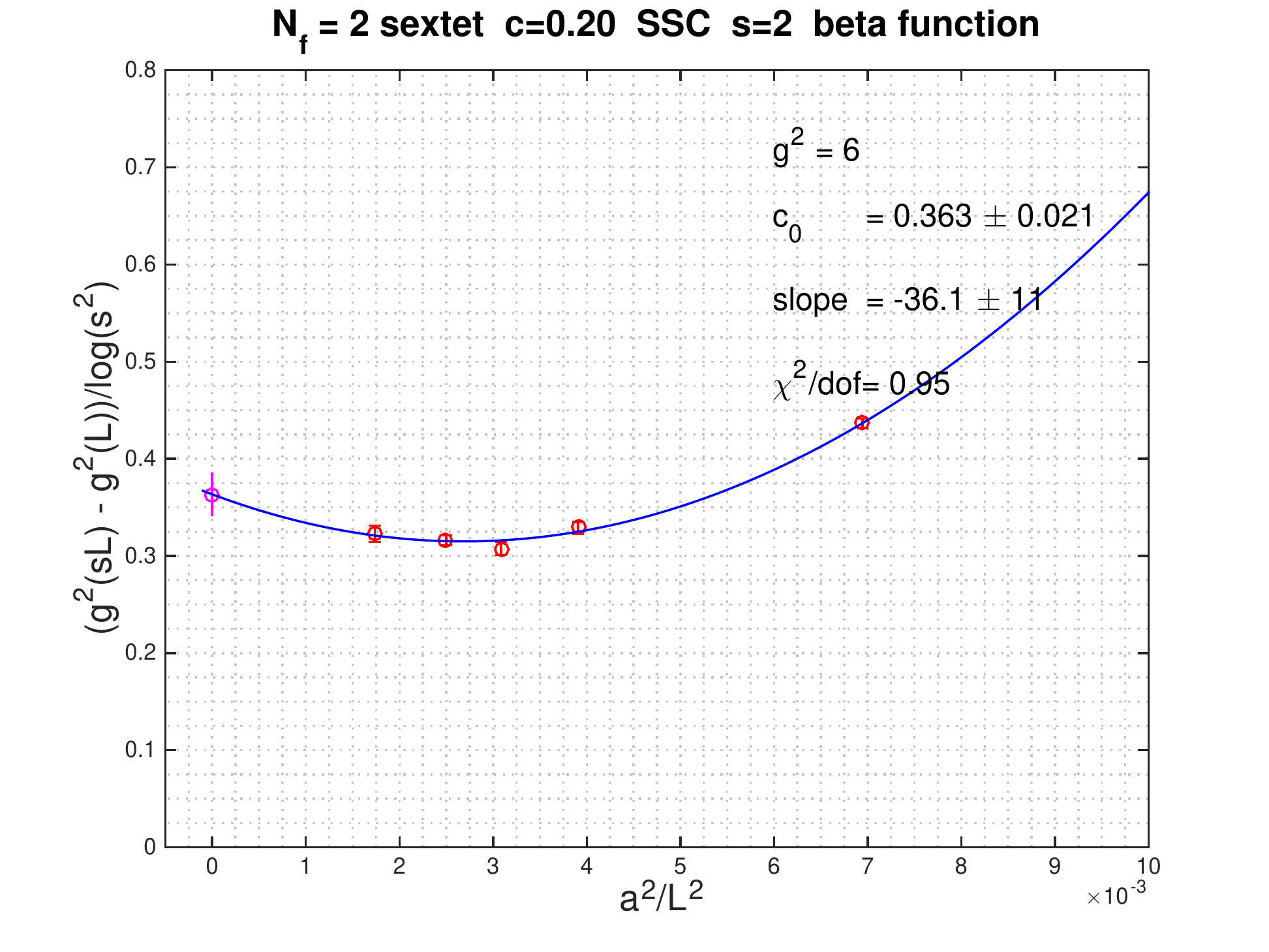}
     \caption{Continuum extrapolations for $N_f = 2$ sextet, making contact with perturbative results at weak coupling and showing consistency with previous analysis at stronger coupling.}
     \label{fig6}
\end{center}
\end{figure}

The $N_f = 2$ sextet model has been our flagship BSM candidate, given the match of 3 Goldstone to 3 Electroweak gauge bosons, compiling evidence of spontaneous chiral symmetry breaking, and a small but non-zero $\beta$-function. A refresh of the $\beta$-function ensembles and analysis is ongoing, we show in Fig.\ref{fig6} examples at weak and strong coupling. Similar to our fundamental rep studies, we find the SSC flow discretization gives accurate results with a controlled continuum extrapolation. At weaker coupling $g^2 = 1$ the non-perturbative $\beta$-function appears to make contact with the perturbative expansion (however the 5-loop $\overline{\rm MS}$ calculation is in a different scheme). At stronger coupling $g^2 = 6$ we see both ${\cal O}(a^2)$ and ${\cal O}(a^4)$ terms being necessary for continuum extrapolation if including $L/a = 12$. The overall behavior is consistent with our published results~\cite{Fodor:2015zna} that the continuum $\beta$-function increases out to $g^2 = 7$, in quatitative difference with the Wilson results of \cite{Hasenfratz:2015ssa}. Our published work presents an RG argument for the validity of staggered fermions, refuting claims repeated in~\cite{Hasenfratz:2017mdh} about the loss of universality in the staggered discretization. Systematic tests of the analysis and its dependence on $c$ and flow discretization are continuing. 

\section{Infinitesimal $\beta$-function}

\begin{figure}[h]
\begin{center}
     \includegraphics[width=.46\textwidth]{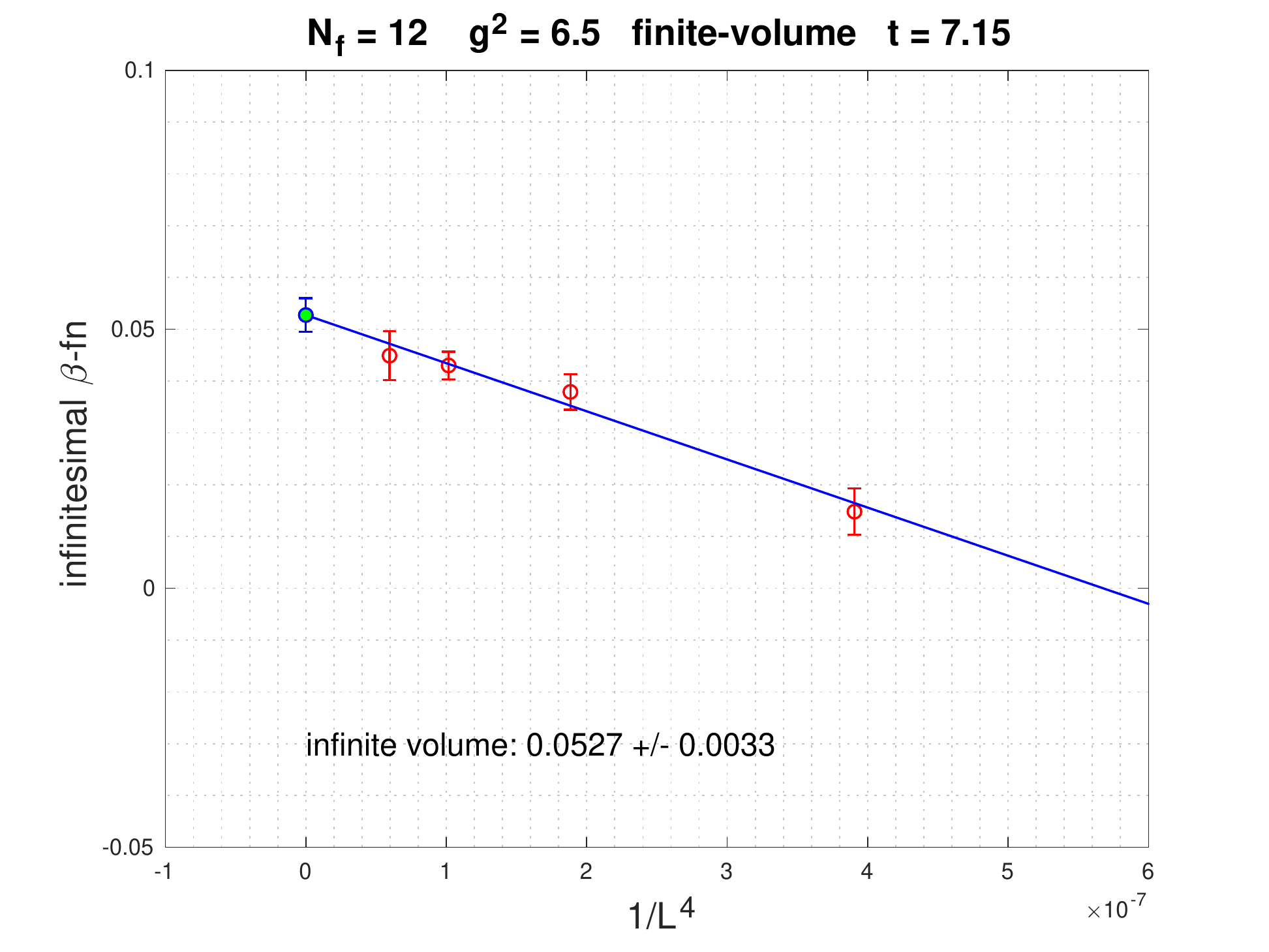}
     \includegraphics[width=.46\textwidth]{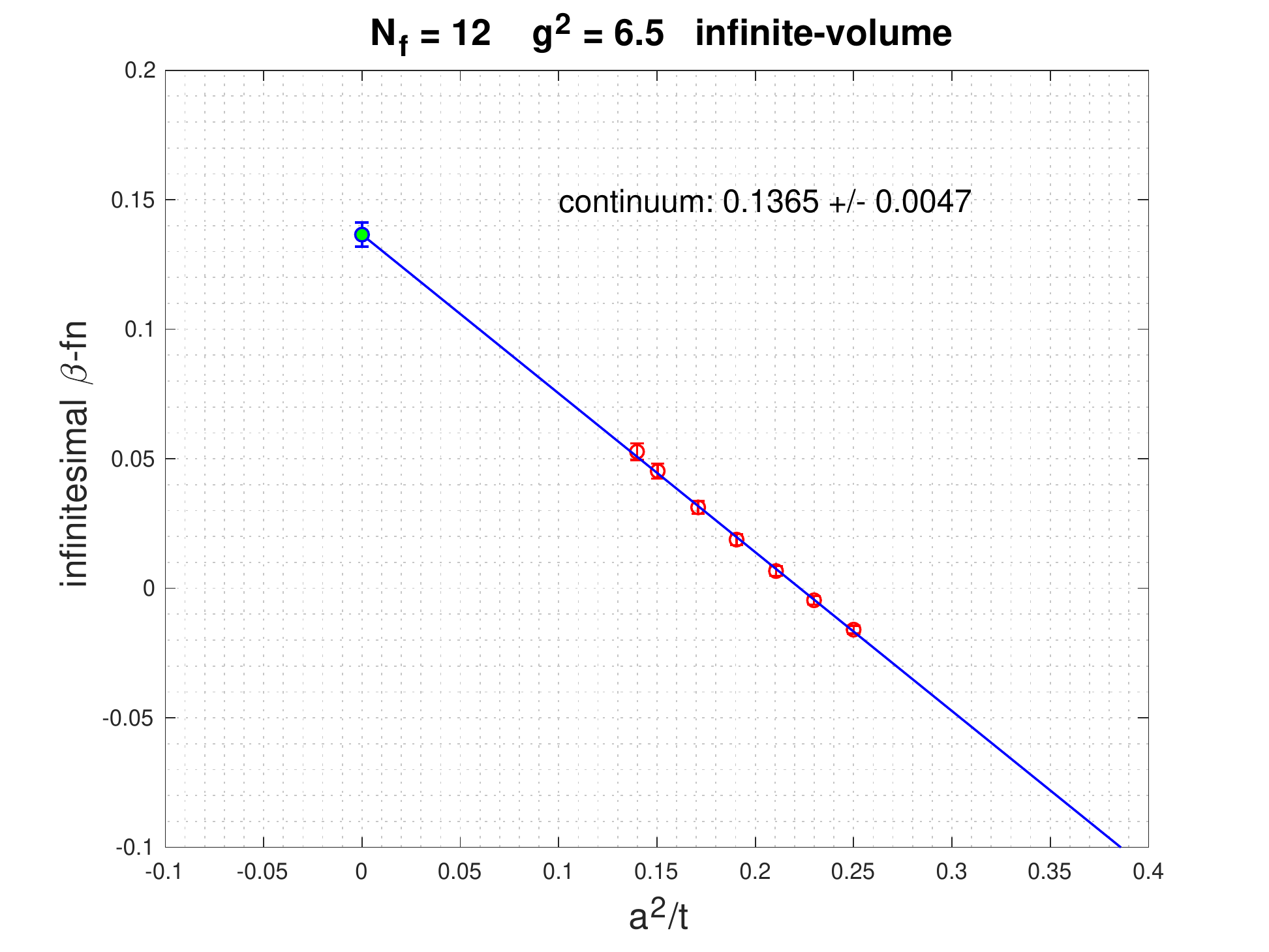}
     \caption{An example of the $N_f = 12$ infinitesimal $\beta$-function. (Left) Simulating directly at zero fermion mass, finite-volume dependence of ${\cal O}(a^4/L^4)$ is removed at fixed $g^2$ and flow time $t$ on volumes $40, 48, 56, 64$, followed by continuum extrapolation of the infinite-volume results at fixed $g^2$ with ${\cal O}(a^2/t)$ (right). The continuum result is in excellent agreement with the finite-step determination in Fig.\ref{fig2}.}
     \label{fig8}
\end{center}
\end{figure}

Since the gradient flow gives the renormalized coupling $g^2(t)$ at any flow time $t$, an infinitesimal $\beta$-function $t \cdot dg^2/dt$ can be measured directly from lattice simulations, which can be extrapolated to infinite volume and the continuum in a controlled fashion. This is an alternative to the finite-volume discrete step-function $[g^2(sL) - g^2(L)]/\log(s^2)$ with scale change $s$. This technique can be applied to the ensembles previously generated for the discrete step-function, hence for a variety of models. We show in Fig.\ref{fig8} an example for the $N_f = 12$ theory. With simulations directly at zero fermion mass, finite-volume dependence of the $L/a = 40, 48, 56, 64$ data is removed as ${\cal O}(a^4/L^4)$ at each choice of reference flow time $t$ where the infinitesimal derivative is measured, approximated with a finite difference accurate to ${\cal O}(\epsilon^4)$. We repeat for a range of $t$ values where the finite-volume effect is under control. The following continuum extrapolation in $a^2/t$ gives an accurate continuum limit in excellent agreement with the discrete step-function results shown in Fig.\ref{fig2}. This independent calculation further bolsters the case that the $N_f = 12$ model is near-conformal with a small but non-zero $\beta$-function. We have also calculated the infinitesimal $\beta$-function for the $N_f = 10$ fundamental rep model as shown in Fig.\ref{fig9}.

\begin{figure}[h]
\begin{center}
     \includegraphics[width=.42\textwidth]{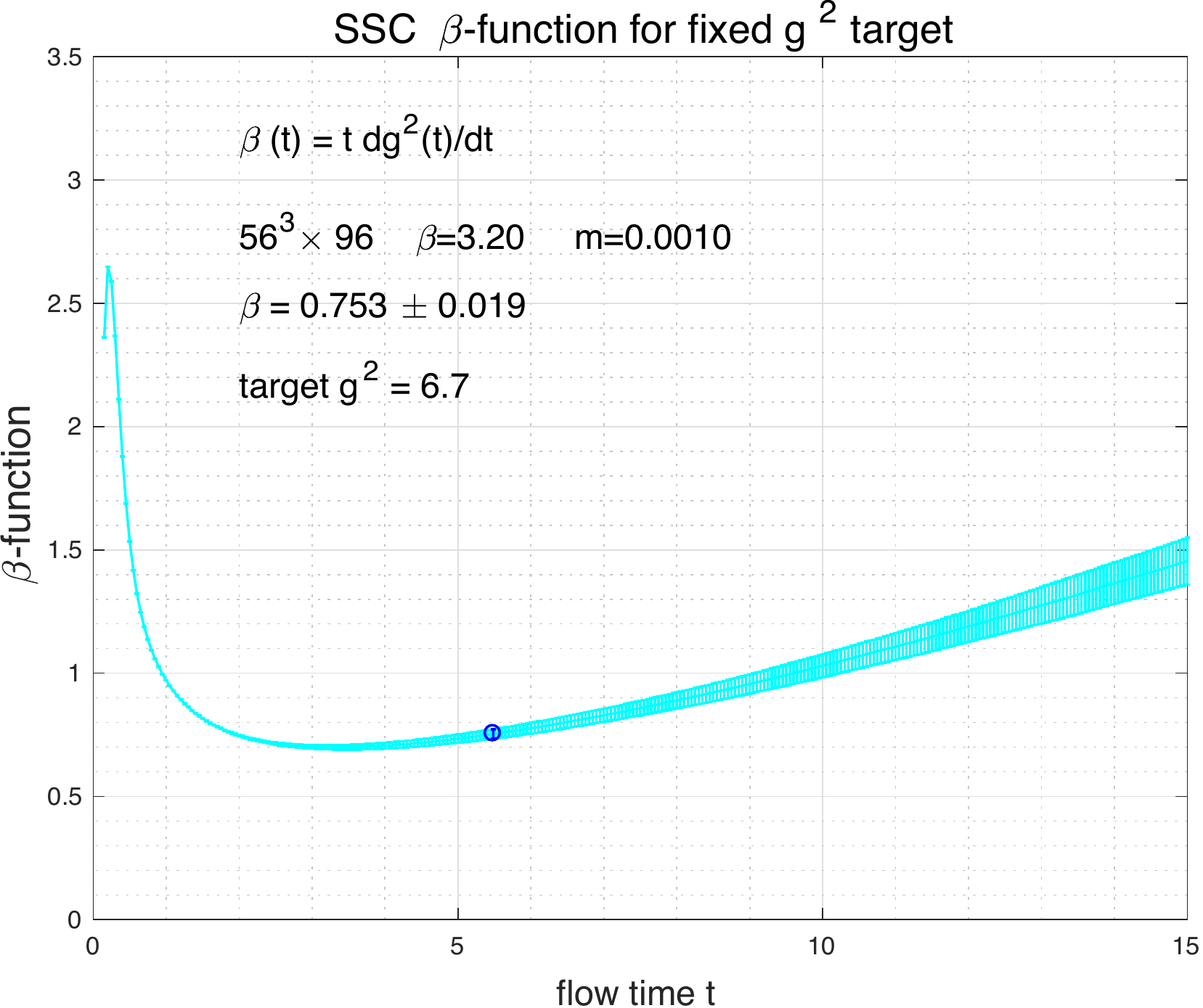}
     \includegraphics[width=.42\textwidth]{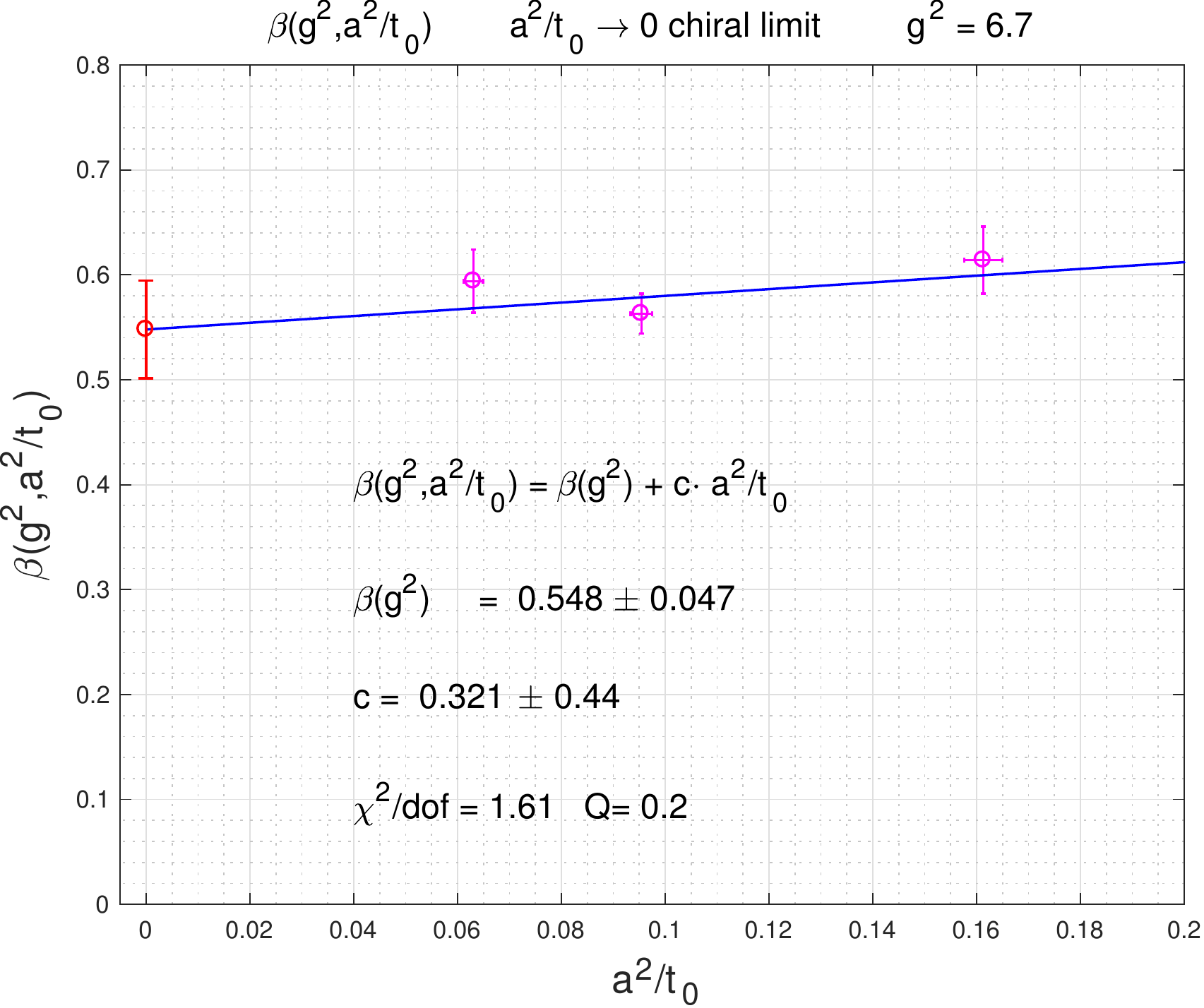}
     \caption{Presented at Lattice 2017, the infinitesimal $\beta$-function with gradient flow for $N_f = 2$ sextet. Finite-volume and mass effects are fitted before ${\cal O}(a^2/t_0)$ continuum extrapolation, with $t_0$ a reference scale.}
     \label{fig7}
\end{center}
\end{figure}

We first presented a calculation of the infinitesimal $\beta$-function at Lattice 2017~\cite{Fodor:2017die} for the $N_f = 2$ sextet model as in Fig.\ref{fig7}. We fitted a set of $p$-regime ensembles to remove finite-volume dependence at finite mass, then took the chiral limit of both the $\beta$-function and a reference scale $t_0$ for each lattice spacing. The final step is a continuum extrapolation in $a^2/t_0$, giving a continuum result at strong coupling matching the fully independent finite-step $\beta$-function measured on zero-mass ensembles. The connection of the $p$-regime and the weak coupling regimes shows the massless theory spontaneously breaks chiral symmetry at strong coupling, developing a mass gap which prevents a conformal phase from emerging. The method was tested for $N_f = 2$ in~\cite{Hasenfratz:2019hpg}. 

\vspace{-6mm}
\acknowledgments
We thank for their support the DOE under grant DE-SC0009919, the NSF under grant 1620845, the NKFIH grant KKP-126769, the Deutsche Forschungsgemeinschaft grant SFB-TR 55, the AEC at the University of Bern, the DOE INCITE program on the ALCF BG/Q platform, USQCD at Fermilab, the University of Wuppertal, and the Juelich Supercomputing Center.

\end{document}